\documentclass[aps,prl,floatfix,twocolumn,showpacs,tightenlines,groupedaddress,superscriptaddress,amsmath]{revtex4}

\usepackage{amssymb}

\usepackage{bm}

\ifx\pdftexversion\undefined
  \usepackage[dvips]{graphicx}
\else
  \usepackage[pdftex]{graphicx}
\fi

\usepackage{color}

\def\tfrac#1#2{{\textstyle\frac{#1}{#2}}}
\newcommand {\half} {\tfrac{1}{2}}

\newcommand{\ket}[1]{\mbox{$|#1\rangle$}}

\newcommand{\ud}{{\uparrow \downarrow}}
\newcommand{\du}{{\downarrow \uparrow}}
\newcommand{\uu}{{\uparrow \uparrow}}
\newcommand{\dd}{{\downarrow \downarrow}}
\newcommand{\da}{{\downarrow}}
\newcommand{\ua}{{\uparrow}}

\newcommand{\Tr}{{\rm Tr}}
\def \be{\begin{equation}}
\def \ee{\end{equation}}
\def \bew{\begin{widetext}\begin{equation}}
\def \eew{\end{equation}\end{widetext}}
\def \bmlett{\begin{mathletters}}
\def \emlett{\end{mathletters}}

\def \nbar{n_{\rm eq} }

\def \NN{{\mathcal N}}

\def \talph{ \tilde{\alpha} }

\def \ua{\uparrow}
\def \da{\downarrow}
\def \ra{\rightarrow}

\def \hrho{\hat{\rho}}

\def \hrho{\hat{\rho}}

\def \hA{\hat{A}}
\def \hD{\hat{D}}

\def \ha{\hat{a}}

\def \hsig{\hat{\sigma}}

\def \omegam{\omega_{\rm M}}
\def \omegaqb{\omega_{\rm qb}}

\begin{document}
\title{Entanglement Dynamics in a Dispersively Coupled Qubit-Oscillator System}
\author{D. Wahyu Utami}
\affiliation{Physics Department, McGill University, Montreal,
Quebec, Canada H3A 2T8}
\author{A. A. Clerk}
\affiliation{Physics Department, McGill University, Montreal,
Quebec, Canada H3A 2T8}

\date{March 4, 2008}

\begin{abstract}
We study entanglement dynamics in a system consisting of a qubit dispersively
coupled to a finite-temperature, dissipative, driven oscillator.  
We show that there are two generic ways
to generate entanglement: one can entangle the qubit either with the phase or the amplitude of the oscillator's motion.  Using an exact solution of the relevant quantum master equation, we study the robustness of both these kinds of entanglement against the effects of dissipation and temperature; in the limit of zero temperature (but finite damping), a simple analytic expression is derived for the logarithmic negativity. We also discuss how the generated entanglement may be detected via dephasing revivals, 
being mindful that revivals can occur even in the absence of any useful entanglement.
Our results have relevance to quantum electromechanics, as well as to circuit QED systems.

\end{abstract}

\maketitle


There exists a long-standing interest in attempting to prepare and detect quantum
states of macroscopic objects or collective degrees of freedom.  Such an experiment would be more than a mere ``proof-of-principle":  it would provide a non-trivial test of our understanding of the  quantum dissipative processes which cause such states to degrade with time, and thus enforce the quantum to classical transition.    Recent advances suggest that submicron-scale mechanical resonators could be excellent candidate systems in which to pursue this goal \cite{SchwabRoukes05}.  Such resonators contain a truly macroscopic number of atoms;  at the same time, they can be fabricated to have both high mechanical frequencies and quality factors.  This suggests that one has some hope of cooling these systems to close to their quantum ground state, and that decoherence due to the dissipative environment of resonator should be slow-- in standard models, decoherence rates scale with the oscillator damping rate \cite{PazHabib93}.  Nanomechanical resonators also have the advantage that they can be strongly coupled to (possibly coherent) electronic degrees of freedom; this has recently been demonstrated to allow sensitive position detection, approaching the fundamental limits set by quantum back-action \cite{Knobel03} \nocite{Naik06, FlowersJacobs07}.

In this Letter, we analyze a promising electromechanical system where a dissipative mechanical resonator is dispersively coupled to a superconducting qubit:  the state of the qubit simply shifts the frequency of the resonator \cite{Wei06, Clerk07a}.  Such a setup has the key advantage of being able to work with qubit states which are first-order insensitive to dephasing due to ever-present charge fluctuations \cite{Blais04, Wallraff04}.  As such, the system is substantially different from the one analyzed in the seminal proposal of  Armour et al.~\cite{Armour02}, which made use of quickly-dephasing superpositions of charge states in the qubit.  While the dispersive coupling allows for longer qubit coherence times, there is a price to pay:  unlike the proposal of Ref.~\onlinecite{Armour02}, the two energy eigenstates of the qubit do not yield different average forces on the oscillator.  As such, generating entanglement is a slightly more involved affair.  
We demonstrate that there are two generic ways to generate non-classical, entangled oscillator-qubit states:  one can entangle the qubit either with the {\em amplitude} of the oscillator's motion, or with its {\em phase}.  
We also study the robustness of these two kinds of entanglement against decoherence due to the dissipative environment, and discuss their detection using coherence revivals.  A fully analytical expression for the entanglement (as measured by the logarithmic negativity) is obtained for the zero
temperature case.  In the finite temperature case, we make use of an exact solution of the master equation to efficiently calculate the time-dependent entanglement.  We stress that the system studied here has already been realized, both with nanomechanical resonators \cite{LaHayePrivateComm}, and with superconducting stripline resonators in circuit QED experiments \cite{Wallraff04}.  Our study also sheds light on general questions of entanglement dynamics in the presence of dissipation, driving and thermal noise.  While entanglement has been studied in a variety of qubit-plus-oscillator models \cite{Costi03, Hines04}, we are not aware of a study involving the dispersive coupling considered here, nor of studies examining the effects of both driving and dissipation.

{\em System.}-- We consider a damped mechanical oscillator 
(frequency $\omegam$) which is dispersively coupled to a qubit (splitting frequency $\omegaqb$).  Setting $\hbar=1$, the Hamiltonian takes the form $H = H_0 + H_{\gamma}$ with:
\begin{eqnarray}
	H_0 &=& 
		 \left(\omegam + \lambda \hsig_z \right) (\ha^\dagger \ha+ \half) + 
	\frac{ \omegaqb}{2} \hsig_z  + f(t) \left(  \ha + \ha^{\dag} \right) 
	\nonumber \\
	\label{eq:HTot}
\end{eqnarray}
Here, $\lambda$ is the strength of the dispersive coupling, $f(t)$ is an external force applied to the resonator, and $H_{\gamma}$ describes the damping of the oscillator (damping rate $\gamma$) by an equilibrium Ohmic bath at temperature $T$.  We stress that such a dispersive coupling can be easily realized in systems having nanomechanical resonators coupled to superconducting qubits, as it emerges naturally from a  
Jaynes-Cumming type coupling in the relevant limit where $\omegaqb \gg \omegam$ \cite{Blais04}.  Dispersive couplings have also been achieved in the same way in recent circuit QED experiments coupling superconducting qubits to stripline resonators \cite{Schuster07}.   

The dispersive coupling implies that the two qubit energy eigenstates $| \ua \rangle$, $| \da \rangle$ each lead to different oscillator frequencies; equivalently, the effective frequency of the qubit depends on the energy of the oscillator.  Unlike previous proposals
\cite{Armour02}, the two qubit states do not yield different oscillator forces, making
entanglement generation somewhat more subtle.  
One can easily show that if the oscillator starts in a thermal state and is not driven, then there is never any qubit-oscillator entanglement:  the dispersive coupling simply leads to a statistical uncertainty in the qubit's frequency.  Thus, to generate non-classical states, one must have a non-equilibrium state where $\langle \ha(t) \rangle$ is non-zero.  To describe this, as well as the effects of dissipation and thermal noise, we will focus on the experimentally relevant regime where $\lambda, \gamma \ll \omega_M$, and use the high-Q form of the Brownian-motion master equation for our system \cite{Clerk07a}:
\begin{eqnarray} 
	\dot{\hrho} &=&
		-i \left [ H_0, \hrho \right ]
	+ \gamma (\nbar+1) \mathcal{D}[\ha] \hrho 
	+ \gamma \nbar
	\mathcal{D}[\ha^\dagger] \rho
	\nonumber \\
	&& + (\Gamma_{\varphi} / 2) \mathcal{D}[ \hsig_z ] \hrho
	\label{eq:Master}
\end{eqnarray}
where for any operator $\hA$ we define 
$	\mathcal{D}[\hA] \hrho = 
		\hA \hrho \hA^{\dag} -  \left(
			\hA^{\dag} \hA \hrho + \hrho \hA^{\dag} \hA
		\right)/2 $.
Here, $\gamma$ is the damping rate of the resonator due the bath, $\nbar$ is a Bose-Einstein factor evaluated at the bath temperature $T$ and energy $\omegam$, and $\Gamma_{\varphi}$ is the intrinsic dephasing rate of the qubit.
We consider the usual situation where the qubit energy relaxation time is much longer than the dephasing time, and ignore $T_1$ processes. 

As shown in Ref.~\cite{Clerk07a}, Eq.~(\ref{eq:Master}) can be solved for arbitrary $T$ by using the fact that an initial Gaussian state remains Gaussian.  With some work, this solution may be written in a physically transparent manner that is especially convenient for entanglement calculations.  For compactness, we focus on the relevant initial condition where at $t=0$ there is no qubit-oscillator entanglement, and the qubit state is $\ket{ \psi} = ( \ket{\ua} + \ket{\da} )/\sqrt{2} $.  Defining the displacement operator $\hD[\alpha] = \exp[\alpha \ha^{\dag} - \alpha^* \ha ]$ and $\hrho_{\sigma \sigma'}$ as the partial trace  $\Tr_{\rm qb}
\left[ \hrho | \sigma' \rangle \langle \sigma| \right]$, the solution takes the form:
\begin{subequations}
\begin{eqnarray}
	\hrho_\uu(t) &=& 
		(1/2) \hD[ \alpha_\uparrow(t)] \cdot \hrho_{\rm eq}[T] \cdot \hD^{\dag}[  \alpha_\uparrow(t)] \\
	\hrho_\dd(t) &=& 
		(1/2) \hD[\alpha_\downarrow(t)] \cdot \hrho_{\rm eq}[T] \cdot \hD^{\dag}[ \alpha_\downarrow(t)] \\
	\hrho_\ud(t) &=&  \left[ \hrho_\du(t) \right] ^{\dag} = 
		(1/2) 
		 e^{ i \omegaqb t } Y(t) \times
			\label{eq:rhoud}	\\
		&&
		\hD[\tilde{\alpha}_\ua(t)] \cdot 
			\left( 
				\hrho_{\rm eq}[T^*(t)] e^{-i \phi(t) (\hat{n}+ \half)}
			\right)
			\cdot \hD^{\dag}[ \tilde{\alpha}_\da(t)] 		
			\nonumber 
\end{eqnarray}
\label{eqs:rho}
\end{subequations}
Here, $\hrho_{\rm eq}(T)$ is  the thermal equilibrium oscillator density matrix at temperature $T$.  As expected, $\hrho_{\uu}$ and $\hrho_{\dd}$ are simply displaced thermal oscillator states, with the displacements $\alpha_{\sigma}(t)$ denoting the conditional means 
$\Big \langle \ha(t)  \cdot |\sigma \rangle \langle \sigma \Big \rangle$.
One has 
	$ \dot{\alpha}_{\ua/\da} = -i \left(
		\omega_M \pm \lambda - i \gamma/2 
	\right) \alpha_{\ua/\da} - i f(t) $.
The dynamics of $\hrho_{\ua \da}$ is more complex.  We see that it also resembles a displaced thermal state.  However, the effective temperature $T^*$ is time dependent,
and the dependence of the effective qubit frequency on the oscillator energy leads to a phase $\phi(t)$.  Defining $\sigma(t) = \coth \left[ (\omegam/ 2T^*) + i \phi/2 \right]$, one finds:
\begin{eqnarray}
	\dot{\sigma} = - \gamma\left( \sigma - (2 \nbar + 1) \right)
		- i \lambda  ( \sigma^2 - 1)
	\label{eq:sigmadot}
\end{eqnarray}
with the initial condition $T^*(0) = T, \phi(0) = 0$.  In addition, the displacements $\talph_{\ua,\da}(t)$ are generally {\em not} the expected, classical displacements $\alpha_{\ua/ \da }(t)$ appearing in $\hrho_{\uu/\dd}$; they only coincide at $T=0$.
One has:
\begin{eqnarray}
	\talph_{\ua / \da} & = &
		\left( a_{\pm} \right)  \pm \frac{ (1 - \textrm{Re} \sigma \mp i \textrm{Im} \sigma ) (a_+ - a_-)}{2 \textrm{Re } \sigma} 
\end{eqnarray}
where we have defined
\begin{eqnarray}
	\dot{a}_{\pm} & = &
		\left( -i 
			\left(\omega_M \pm \lambda  \textrm{Re }\sigma   \right) - 
				\frac{\tilde{\gamma}}{2} \right) a_{\pm} 
			- i f(t)
		\label{eq:adot}	
\end{eqnarray}
and $\tilde{\gamma} = \gamma + \lambda \omega_M \textrm{Im } \sigma$.  For $T=0$, $\sigma \ra 1$, and $\talph_{\ua/\da}(t) = a_{+/-}(t) = \alpha_{\ua/\da}(t)$.
Finally, the parameter $Y(t)$ describes the fact that due to the dissipative bath, the qubit-oscillator system will generally not be in a pure state
($p_n(T) =  \langle n| \hrho_{\rm eq}(T) |n \rangle$):
\begin{eqnarray}
	Y(t)& = &\frac{ e^{ - \Gamma_{\varphi} t}
		\exp \left[
			-  i \lambda \int_0^t dt'  \left(\sigma +
				2 a_+ a_-^* \right) \right]  }
	{
		 \sum_n p_n(T^*) e^{-i \phi (n + 1/2)}
			\langle n | \hD^{\dag}[ \tilde{\alpha}_\da] 
				\hD[\tilde{\alpha}_\ua] | n \rangle }
	\label{eq:Y}
\end{eqnarray}

{\em Entanglement}-- 
Similar to many recent works, we use the logarithmic negativity $E_N$ to quantify the amount of qubit-oscillator entanglement in our system \cite{Horodecki98,Vidal02}.  $E_N$ is a rigorous entanglement monotone applicable to mixed state systems, and is strictly zero for unentangled systems.  One has $E_N=\log_2(2  \mathcal{N}  +1)$, where the negativity $\mathcal{N}$ is the absolute value of the sum of the negative eigenvalues of the partially-transposed density matrix $\hrho^{PT}$.  

To create non-classical, entangled qubit-oscillator states, we focus on the initial
condition described above Eqs.~(\ref{eqs:rho}):  at $t=0$
there is no qubit-oscillator entanglement, and the qubit has been prepared (e.g.~via a $\pi/2$ pulse) in an equal superposition of its two eigenstates.  The two qubit states in this superposition each lead (via the dispersive coupling) to different oscillator frequencies; this difference will be exploited to yield entangled states which, in the absence of  dissipation or thermal noise, would have the simple form
\begin{eqnarray}
	\ket{\psi_{\rm tot}(t)} \propto \left[
			 \ket{\ua} \ket{\alpha_{\ua}(t) } + 
			 e^{-i \omegaqb t} \ket{\da} \ket{\alpha_{\da}(t)}
	\right]
\end{eqnarray}
There are thus two generic ways to generate entanglement.  The first is to entangle the qubit with the {\em amplitude} of the oscillator's motion, i.e~$| \langle \ha \rangle |$.
The second generic approach is to entangle the qubit with the {\em phase} of the resonator, i.e.~$\arg \langle \ha \rangle$.  

In the zero-temperaure limit (but still with $\gamma > 0$), the solution of the master equation in Eqs.~(\ref{eqs:rho}) yields an exact expression for $E_N(t)$.  Letting $\cos \left[ \theta(t) \right] = | \langle \alpha_{\ua}(t) | \alpha_{\da}(t) \rangle |$, 
one finds:
\begin{eqnarray}
	\mathcal{N}(t)=\frac{1}{4}
		\left[ \sqrt{ (1-|Y| )^2+4 |Y|\sin^2\theta } - \left(1-|Y| \right) \right].
\label{eq:ZeroTNegativity}
\end{eqnarray}
where for $T=0$, Eq.~(\ref{eq:Y}) for $Y$ yields:
\begin{eqnarray}
|Y(t)| &=& \frac{ 
		e^{-\Gamma_{\varphi} t }
		\exp \left[ - 2 \lambda \int_0^t dt'
		\left( 
			\left| 
				\alpha_{\ua} \alpha_{\da}
			\right| 
				\sin \phi_{\ua \da}  
			\right) \right]
		} {\cos \theta(t) }		
	\label{eq:purity}
\end{eqnarray}
with $\phi_{\ua \da}(t) = -\arg \left[ \alpha_{\ua}(t) \alpha_{\da}^*(t) \right]$.  As expected, $\NN$ (and hence the entanglement) is an increasing function of the distinguishability $\sin^2 \theta$ of the two oscillator states.  The factor $|Y(t)|$ describes as before the impurity of the oscillator-qubit system at times $t > 0$:  as time progresses, the bath
can distinguish the two oscillator states $\ket{ \alpha_{\sigma} }$, 
thus reducing the oscillator-qubit entanglement.  One might expect that $|Y(t)|$ should only depend on the history of the overlap $\cos \theta(t')$; this is not the case.  Instead, the decay of $|Y(t)|$ is sensitive to the history of $\sin \phi_{\ua \da}(t)$, i.e. the sine of the relative phase between the two oscillator coherent state amplitudes.  Somewhat surprisingly, times when $\phi_{\ud} = \pi$ do not contribute to the decay of $|Y(t)|$, even though the overlap between the two oscillator states at such times is maximally small.

In the case where $T > 0$, we again use our exact Eqs.~(\ref{eqs:rho}) - 
(\ref{eq:adot}) to solve for the system dynamics.  The form of $\hrho(t)$ given in Eqs.~(\ref{eqs:rho})
then allows for a simple numerical evaluation of $\NN$:  one simply converts $\hrho$ into a matrix in a basis of of displaced Fock states, and then numerically finds the partial transpose and the corresponding negative eigenvalues.  For $T>0$, $E_N(t)$ is not simply a function of the
overlap $\langle \alpha_{\ua} | \alpha_{\da} \rangle (t)$ and the purity $Y(t)$ (as given by the full expression Eq.~(\ref{eq:Y}))-- the entire matrix structure of $\hrho_{\ua \da}$ is relevant. 

\begin{figure}[t] 
	\centering
	\includegraphics[width=8cm]{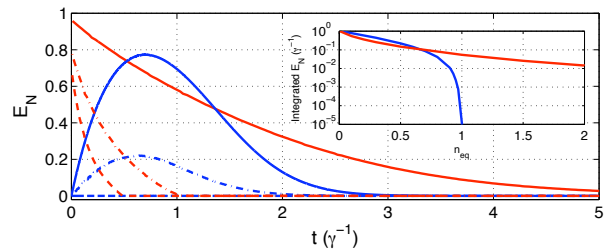}
	\caption{
	Log negativity $E_N$ versus time.
	Blue lines correspond to the amplitude-entanglement setup, red lines correspond to 
	the envelope of the oscillating $E_N$ in the phase-entanglement setup.  Solid lines are for $\nbar = 0$, dot-dashed for
	$\nbar = 0.5$, dashed for $\nbar =1.0$. For the phase-entanglement curves, we 
	took $\alpha_0=0.76$ to maximize the total integral of $E_N(t)$ at 
	$\nbar=0$.  We also chose $\alpha_f=3.74$ so that the integral of 
	$E_N(t)$ for amplitude entanglement at $\nbar = 0$ is the same as the phase case. 
	The inset shows
	the total time integral of $E_N$ for both phase and amplitude 
	entanglement as a function of $\nbar$.
	In all cases, $\lambda = 
	0.01 \omegam, \gamma = 10^{-5} \omegam, \Gamma_{\varphi}=0$.    }
	\label{fig:AmpEnt}
\end{figure}

\begin{figure}[h]
\centering
\includegraphics[width=8cm]{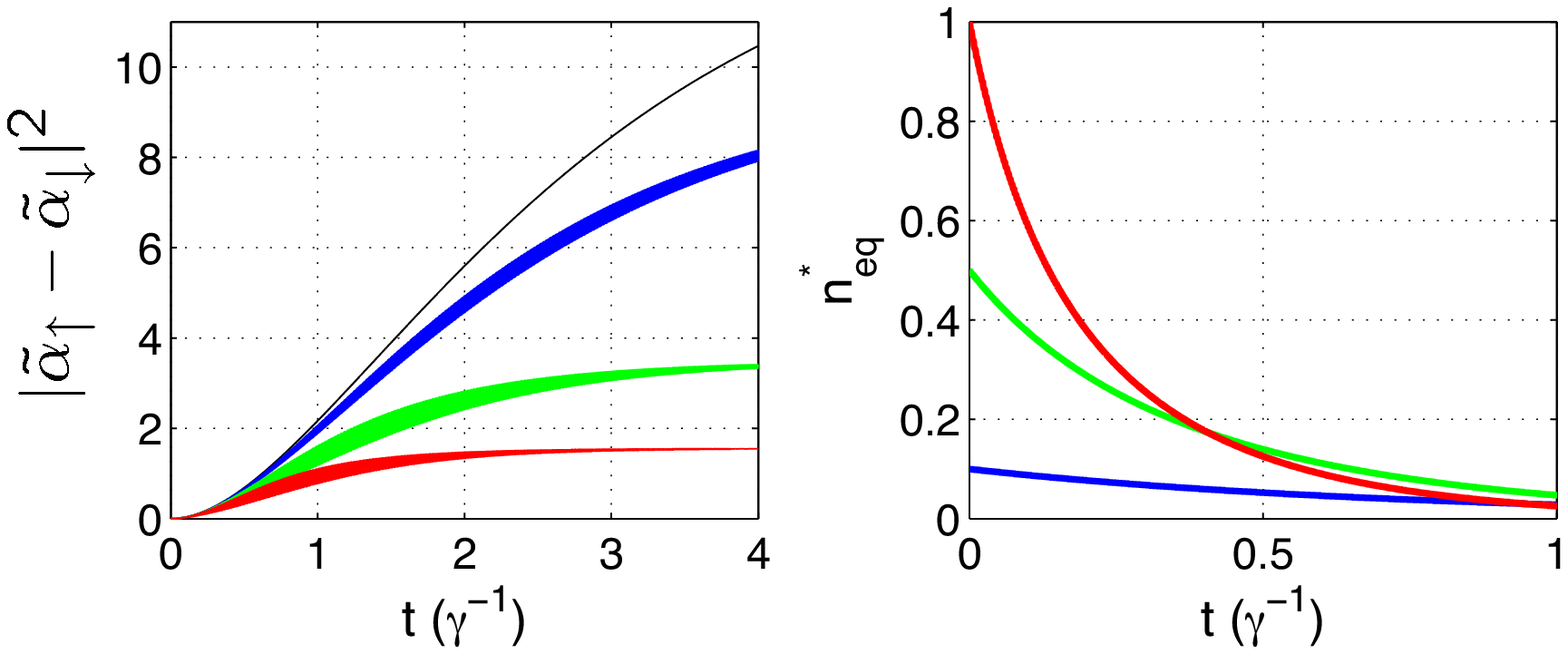}
\caption{
Left: Distance between effective oscillator displacements
$\talph_{\ua}(t)$,$\talph_{\da}(t)$ entering the expression for $\hrho_{\ud}$.
The apparent width of some curves is due to rapid oscillations with frequency
$\lambda / \pi$.
Right: Bose-Einstein factor associated with the effective temperature $T^*(t)$ appearing
in the expression for $\hrho_{\ud}$.  Both cases are for the amplitude entanglement setup with $\alpha_f = 3.74$.  Temperatures are $\nbar = 0$ (black), $\nbar = 0.1$ (blue),$\nbar = 0.5$ (green), $\nbar = 0.1$ (red).
$\lambda, \gamma$ and  $\Gamma_{\varphi}$ as in Fig. 1.  For $T>0$, one finds $\talph_\sigma \neq \alpha_\sigma$; temperature reduces the distinguishability of the $\talph_\sigma$.  
}
\label{fig:AlphaTilde}
\end{figure}

{\em Amplitude entanglement}--   To entangle the qubit with the amplitude of the oscillator's motion, we start at $t=0$ with the qubit in a superposition of its eigenstates, and the oscillator in a thermal state with
$\langle \ha \rangle = 0$.  The oscillator is then driven with a force $f(t) = \gamma \alpha_{f} \cos[ (\omegam + \lambda) t ]$ ($\alpha_{f} > 0$).  In the relevant limit of a high-Q oscillator where $\gamma \ll \lambda$, $f(t)$ will cause 
$| \alpha_{\ua}(t) |$ to grow to a large value $\alpha_{f}$, while $|\alpha_{\da}|$ will be smaller by a large factor $ \omegam / \gamma$:  hence, the amplitude of the oscillator's motion will become entangled with state of the qubit, and we would expect the qubit-oscillator entanglement to grow with time.  However, at long enough times, the dissipative bath coupled to the oscillator will destroy this entanglement:  the bath can distinguish the two states $\ket{ \alpha_{\sigma} }$, as described by $Y(t)$.
These two competing tendencies lead to $E_N$ being a non-monotonic function of time;
this is shown in Fig.~\ref{fig:AmpEnt}.  Note that at $T=0$, one can 
use Eqs.~(\ref{eqs:rho}) and (\ref{eq:ZeroTNegativity}) to derive simple, exact
expressions for $E_N(t)$ for the amplitude entanglement setup; these will be presented
elsewhere.  


Fig.~\ref{fig:AmpEnt} also demonstrates that amplitude entanglement is
dramatically suppressed by finite temperature.  On a heuristic level, even though the two amplitudes $\alpha_{\ua}, \alpha_{\da}$ continue to have very different magnitudes at $T>0$ (they are of course independent of $T$), the displacements $\talph_{\ua}, \talph_{\da}$ which determine $\hrho_{\ud}$ (c.f. Eq.~(\ref{eq:rhoud}))become less distinguishable as $T$ is increased.  This behaviour is shown in Fig.~\ref{fig:AlphaTilde}.



{\em Phase Entanglement}-- 
To entangle the qubit with the resonator phase, we prepare the system at $t=0$ so that the qubit is again in a superposition of its two eigenstates, and the oscillator is in a state of motion characterized by the coherent state amplitude $\alpha_0$ (e.g.~one could drive the oscillator at $\omegam$, keeping the coupling to the qubit off until $t=0$).  At $t=0$, we turn off all driving forces on the oscillator and let the coupled system evolve.  The magnitudes of both coherent states $\alpha_{\ua}, \alpha_{\da}$ will be identical,  and will decay at a rate $\gamma$.  In contrast, the {\it phase} of the oscillator coherent state will wind at a frequency determined by the qubit.  We have thus prepared a state where the phase, not the amplitude, of the oscillator's motion is entangled with the qubit.  

\begin{figure}
\centering
\includegraphics[width=8cm]{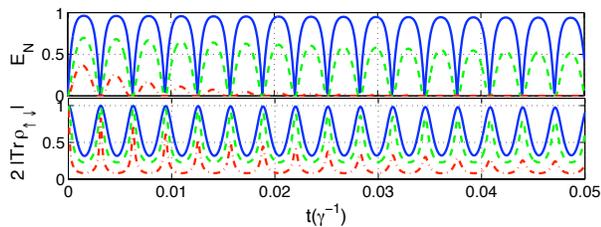}
\caption{Upper panel: time-dependent log negativity $E_N$ for the phase entanglement setup.    Dashed line is
for $\nbar = 0$, dashed-dotted line for $\nbar = 1.0$, solid for $\nbar = 5.0$.
Bottom panel:  Qubit recoherences associated with phase entanglement, same parameters.  
$\alpha_0 = 0.76$; $\lambda, \gamma, \Gamma_{\varphi}$ as in Fig. 1.}
\label{fig:PhaseEnt}
\end{figure}

At zero temperature, one can again obtain simple analytic expressions from
Eq.~(\ref{eq:ZeroTNegativity}); these will be presented elsewhere.
Shown in the top panel of Fig.\ref{fig:PhaseEnt} is $E_N(t)$ for the phase entanglement setup.  The entanglement drops to zero periodically at a frequency $\lambda / \pi$: at these times, the phases of the two oscillator states 
$\alpha_{\ua}, \alpha_{\da}$ are aligned,
implying $\cos \theta = 1$.  This alignment also implies that the coherence of the qubit is
restored, as is shown in the bottom panel of Fig.~\ref{fig:PhaseEnt}, where we plot $2 | \Tr \hrho_{\ud} |$.  

\begin{figure}
\centering
\includegraphics[width=8cm]{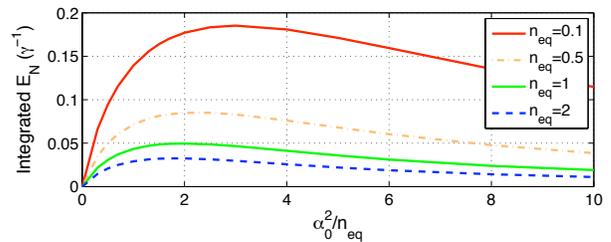}
\caption{Total entanglement (as measured by the time integral of $E_N$) for the phase entanglement setup, as a function of the initial coherent state amplitude $\alpha_0$ and $\nbar$.  For a finite $\nbar$, there is an optimal value of $\alpha_0$ which maximizes the total entanglement.  In all cases, $\lambda = 0.01 \omegam, \gamma = 10^{-5} \omegam, \Gamma_{\varphi}=0$.
}
\label{fig:IntPhaseEnt}
\end{figure}

It is interesting to compare the phase and amplitude entanglement setups; this is done in Fig.~\ref{fig:AmpEnt}, where we plot only the enveloped of the oscillating entanglement in the phase case.
As can be seen from the inset, phase entanglement is more robust against non-zero $T$ than amplitude
entanglement; the total integral of $E_N(t)$ decays far more slowly as a function of $\nbar$ in the phase case, for parameters that yield the same entanglement at $\nbar = 0$.  Further, while both kinds of entanglement are suppressed by non-zero $T$, entanglement in the phase case can remain large for short times (i.e.~for times $t$ such that $1/ \lambda \ll t \ll 1/\gamma$).  Thus, phase entanglement shows a certain increased resilience against thermal dissipation compared to amplitude entanglement.
For a fixed bath temperature, there is in general an optimal value of $\alpha_0$ which maximizes phase entanglement; this is shown in Fig.~\ref{fig:IntPhaseEnt}.  For $\alpha_0 = 0$, $E_N(t)=0$ for all times, where as for too large an $\alpha_0$, the bath very rapidly distinguishes the two oscillator states in the superposition.

\begin{figure}[b]
\centering
\includegraphics[width=8cm]{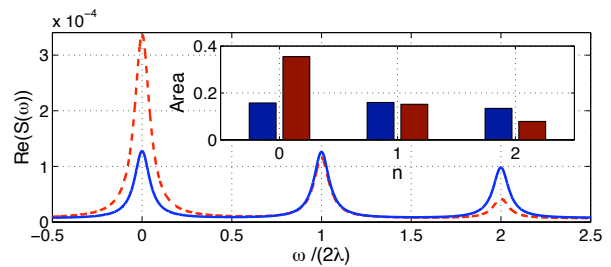}
\caption{$S(\omega)$, the real part of the Fourier transform of the qubit's time-dependent coherence 
2 $| \Tr \hrho_{\ud} (t) |$, for the phase-entanglement setup. The red-dashed curve is for $\nbar = 0.5$, $\alpha_0 = 0$.  There are revivals of coherence, but strictly zero qubit-oscillator entanglement; $S(\omega)$ show peaks whose area follows a geometric series.  In contrast, the solid blue curve corresponds to $\alpha_0 = 1.23$, $\nbar = 0.5$.  Here, one gets both coherence revivals {\em and} entanglement.  The dephasing spectrum is markedly different: the peak areas {\em do not} form a geometric series.  The inset shows the integrated area for each spectra in the two cases.
$\lambda = 0.01 \omegam$, $\gamma = 10^{-5} \omegam$ and $\Gamma_{\varphi} = 10^{-3} \omegam$.}
\label{fig:Ph_newspectrum.eps}
\end{figure}

{\em Detecting Entanglement }-- We now turn to the detection of phase entanglement using coherence revivals.  As shown in Fig.~\ref{fig:PhaseEnt}, such entanglement leads to qubit recoherences:  the magnitude of the qubit's off-diagonal density matrix element $|\Tr \hrho_{\ud}(t)|$ is non-monotonic in time.  This quantity represents the time-dependent dephasing of the qubit, and is measurable via either a standard Ramsey interference experiment \cite{Ithier05}, or via state tomography.  While such revivals of coherence have been used to detect non-classical states in other situations \cite{Haroche05}, and have been proposed as a way to detect entanglement in NEMS \cite{Armour02}, one must be careful:  it is possible to have coherence revivals without any qubit-oscillator entanglement.  In our system, a purely thermal state oscillator state with $\langle \ha(t) \rangle = 0$ yields dephasing revivals, but zero qubit-oscillator entanglement.

Despite this caveat, one can still use dephasing revivals as a proxy for detecting entanglement.  The basic idea is that the Fourier spectrum of the time-dependent dephasing lets one unambiguously distinguish an initial thermal state with $\alpha = 0$ and $E_N(t) = 0$, from a phase-entangled state where $\alpha \neq 0$.  This difference is essentially the number splitting effect discussed in 
Refs.~\cite{Dykman87, Gambetta06, Clerk07a}, 
and measured in Ref.~\cite{Schuster07}:  the peaks in the Fourier spectra are directly related to the number distribution in the oscillator.  Thus, in a thermal state, one expects the peaks to follow a simple geometric series.  In contrast, when $\alpha_0$ is non-zero, the distribution begins to resemble more the Poisson distribution associated with a number state.  Observing this difference, and comparing it to theory, would be a convincing way to detect the phase entanglement we have described.  The difference between the two spectra is shown in Fig.~\ref{fig:Ph_newspectrum.eps}.



{\em Conclusion}--  We have studied entanglement in a dispersive qubit - oscillator system, identifying two generic kinds of entanglement: phase and amplitude entanglement.  We have shown that in general, phase entanglement is more robust against the effects of dissipation, and have shown how it may be detected through the analysis of revivals in the qubit's coherence.

%

%



\bibliographystyle{apsrev}
\bibliography{ACRefs}


\end{document}